\documentclass[useAMS]{mn2e}

\usepackage{ulem} 
\usepackage{graphicx}
\usepackage{ulem}
\usepackage{rotating}

\def\lapprox{\hbox{\lower .8ex\hbox{$\,\buildrel < \over\sim\,$}}}
\def\gapprox{\hbox{\lower .8ex\hbox{$\,\buildrel > \over\sim\,$}}}

\title[\textit{HST\/} observations of the WD CS in NGC\,6819]{
\textit{Hubble Space Telescope} observations of the
  \textit{Kepler}-field cluster NGC\,6819. I. The bottom of the white dwarf
  cooling sequence\thanks{
Based on observations with  the NASA/ESA {\it Hubble Space Telescope},
obtained at  the Space Telescope Science Institute,  which is operated
by  AURA, Inc.,  under NASA  contract NAS  5-26555, under  GO-11688 \&
GO-12669.
}
}
\author[L.\ R.\ Bedin et al.]{ 
L.\ R.\ Bedin$^{1}$\thanks{E-mail: luigi.bedin@oapd.inaf.it}, 
  M.\ Salaris$^{2}$, 
  J.\ Anderson$^{3}$, 
  S.\ Cassisi$^{4}$, 
  A.\ P. Milone$^{5}$, 
  G.\ Piotto$^{6,1}$,  
\newauthor
  I.\ R.\ King$^{7}$, and 
  P.\ Bergeron$^{8}$. 
\\  
$^{1}$INAF-Osservatorio Astronomico di Padova, Vicolo dell'Osservatorio 5, I-35122 Padova, Italy\\
$^{2}$Astrophysics Research Institute, Liverpool John Moores University, 146 Brownlow Hill, Liverpool L3 5RF, UK\\
$^{3}$Space Telescope Science Institute, 3800 San Martin Drive, Baltimore, MD 21218\\
$^{4}$INAF-Osservatorio Astronomico di Collurania, via M. Maggini, 64100 Teramo, Italy\\
$^{5}$Research School of Astronomy and Astrophysics, The Australian National University, Cotter Road, Weston, ACT, 2611, Australia\\
$^{6}$Dipartimento di Fisica e Astronomia ``Galileo Galilei'', Universit\`a di Padova, Vicolo dell'Osservatorio 3, I-35122 Padova, Italy\\
$^{7}$Department of Astronomy, University of Washington, Box 351580, Seattle, WA 98195-1580\\
$^{8}$D\'epartement de Physique, Universit\'e de Montr\'eal, C.P. 6128, Succ. Centre-Ville, Montr\'eal, Qu\'ebec H3C 3J7, Canada\\
}
\begin{document}

\date{Accepted 2015 January 12.  Received 2015 January 9; in original form 2014 November 27}
\pagerange{\pageref{firstpage}--\pageref{lastpage}} \pubyear{2015}

\maketitle

\label{firstpage}

\begin{abstract}
We use  \textit{Hubble Space Telescope} (\textit{HST\/})  to reach the
end  of   the  white   dwarf  (WD)  cooling   sequence  (CS)   in  the
solar-metallicity   open  cluster   NGC\,6819.   Our  photometry   and
completeness tests show a sharp drop in the number of WDs along the CS
at magnitudes  fainter than $m_{\rm  F606W} = 26.050 \pm  0.075$. This
implies  an  age of  $2.25\pm0.20$~Gyr,  consistent  with  the age  of
$2.25\pm0.30$~Gyr obtained from fits to the main-sequence turn-off.
The  use  of  different   WD  cooling  models  and  initial-final-mass
relations have  a minor impact  the WD age  estimate, at the  level of
$\sim$0.1 Gyr.\\
As an important  by-product of this investigation we  also release, in
electronic format, both the catalogue  of all the detected sources and
the atlases of the region (in two filters).  Indeed, this patch of sky
studied  by \textit{HST}  (of  size $\sim$70  arcmin$^2$) is  entirely
within the main $Kepler$-mission  field, so the high-resolution images
and deep catalogues will be particularly useful.
\end{abstract}


%
\section{Introduction}
\label{introduction}
%

During the last few decades both observations and theory have improved
to a level that has made  it possible to employ white dwarf (WD) stars
for estimating ages of  stellar populations in the solar neighbourhood
(i.e., Winget  et al.\  1987; Garc\'ia-Berro et  al.\ 1988;  Oswalt et
al.\ 1996), open (i.e., Richer et al.\ 1998; von Hippel~2005; Bedin et
al.\ 2008a, 2010) and globular (i.e., Hansen et al.\ 2004, 2007; Bedin
et al.\ 2009) clusters.

Methods  to determine stellar  population ages  from their  WD cooling
sequences  are usually  based on  the  comparison of  the observed  WD
luminosity function  (LF --  star counts as  a function  of magnitude)
with theoretical ones calculated from WD isochrones.
When considering star  clusters, owing to the single  (and finite) age
of  their  stars,  the   more  massive  WDs  formed  from  higher-mass
short-lived progenitors pile up at  the bottom of the cooling sequence
(CS), producing a turn to the blue (a turn towards lower radii) in the
isochrones.
At   old  ages,   when  the   WD  ${\rm   T_{eff}}$   decreases  below
$\approx$5000~K, the  contribution by collision-induced  absorption of
molecular  hydrogen (Hansen 1998)  to the  opacity in  the atmospheres
reduces  the   infrared  flux  and   increase  the  flux   at  shorter
wavelengths.   This produces  a turn  to the  blue of  the  colours of
individual  cooling tracks, that  enhances the  blue excursion  at the
bottom of old WD isochrones.
The existence  of a  minimum WD luminosity  due to the  cluster finite
age, together with  the accumulation of WDs of  different masses and a
general increase  of WD cooling  times with decreasing  luminosity (at
least before  the onset of Debye  cooling) translates into  a peak and
cut-off in the LF.
Comparisons of observed and predicted absolute magnitudes of the WD LF
cut-off provides the population age.

The  discovery  of  a second,  brighter  peak  in  the  WD LF  of  the
metal-rich  open cluster NGC\,6791  (see Bedin  et al.\  2005a, 2008a,
2008b  for  the discovery  and  possible  interpretations) has  raised
questions about  our understanding  of the CS  in simple  systems like
open clusters, and their use for age dating of stellar populations.
In particular,  this bright  peak has been  interpreted by  Kalirai et
al.\  (2007) as due  to a  population of  massive He-core  WDs, whilst
Bedin  et  al.\  (2008b) have  explained  it  in  terms of  a  sizable
population of WD+WD  binaries.  As for the fainter  peak --expected to
be the  {\sl real} age indicator--  the age obtained  from standard WD
models is in  conflict ($\sim$ 2\,Gyr younger) with  that derived from
the  cluster  main-sequence (MS)  turn-off  (TO),  and  the age  later
obtained from  the cluster eclipsing  binaries studied by  Brogaard et
al.\ (2012).  This  discrepancy has led to a  detailed reevaluation of
the  effect of  diffusion of  ${\rm ^{22}Ne}$  in the  CO  core before
crystallization  sets  in  (e.g.,   Bravo  et  al.\  1992,  Deloye  \&
Bildsten~2002). As  shown by Garc\'ia-Berro et al.\  (2010, 2011) with
full evolutionary calculations, at the old age and high metallicity of
this  cluster (about  twice  solar), the  extra-energy contributed  by
Ne-diffusion in the liquid  phase slows down substantially the cooling
of the models and can bring into agreement WD, TO and eclipsing binary
ages.

This result highlights very clearly the need for further observations,
and the importance of studying WD ages in comparison with TO estimates
in individual clusters.
As   WDs  lie   in  one   of   the  least-explored   regions  of   the
colour-magnitude diagram (CMD), we are carrying out a campaign to find
out whether the case of  NGC\,6791 is unique or whether other clusters
with similar WD CSs might exist.
Our purpose is to extend our  knowledge of the dependence of WD LFs on
cluster age and metallicity.
So far we  have investigated two other open  clusters:\ NGC\,2158 from
space  (Bedin et al.\  2010), and  M\,67 from  the ground  (Bellini et
al.\ 2010); both of them show canonical WD CSs (hence LFs).

The aim  of the present  work is to  investigate the WD CS  of another
open   cluster,  NGC\,6819,  that   is  within   the  $Kepler$-mission
field. NGC\,6819  has solar metallicity  (Bragaglia et al.\  2001), is
about a fourth as old  as NGC\,6791 (Anthony-Twarog et al.\ 2013), and
somewhat less  massive (as  can be inferred  from their images  in the
Digital Sky Survey).

Section~2  will describe  our  observations and  WD selection,  whilst
Section~3 presents the theoretical  analysis of the WD LF.  Sections~4
and 5  discuss our proper  motion analysis and present  the electronic
material we make publicly available.  Conclusions close the paper.

%

%
\section{Observations, Measurements and Selections}
%

All data presented here  were collected with two different instruments
at the focus  of the {\it Hubble Space  Telescope} ({\it HST\/}) under
two programs taken at different epochs, GO-11688 and GO-12669 (on both
PI:\ Bedin).

\begin{figure*}
\begin{center}
\includegraphics[width=178mm]{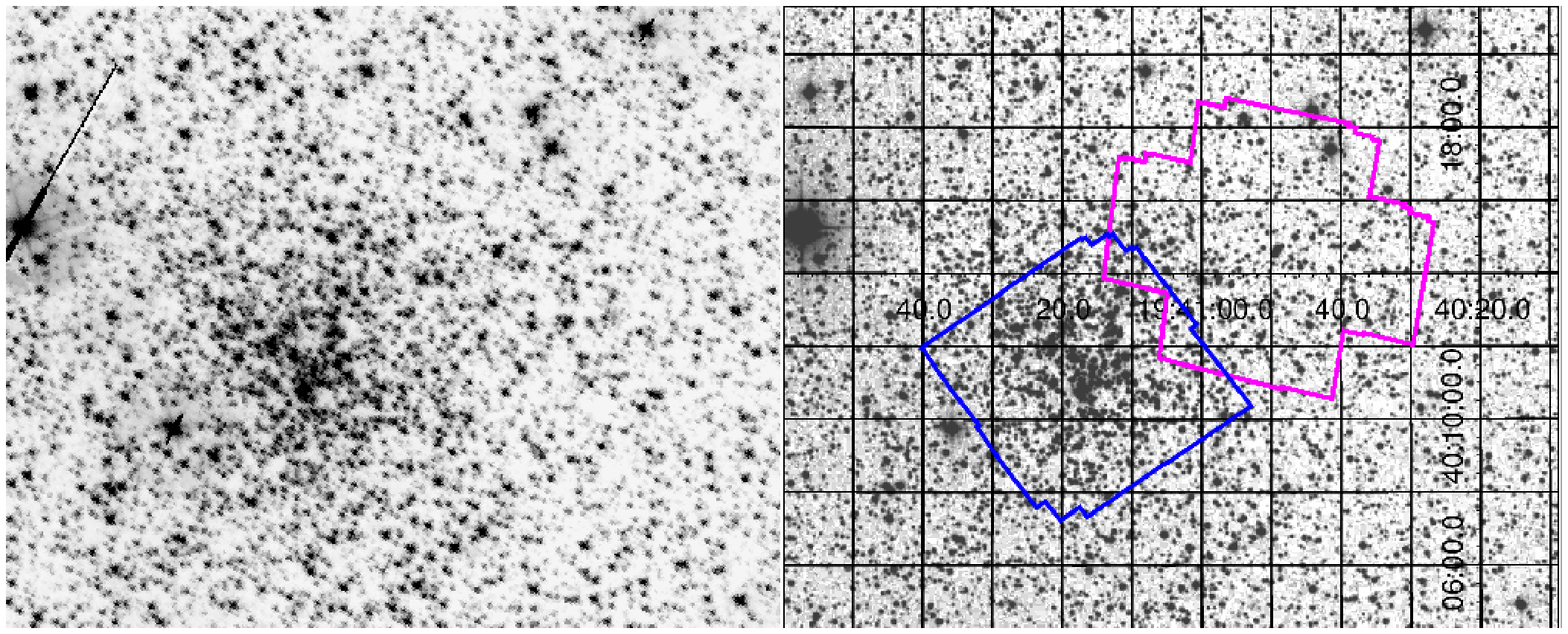}
\includegraphics[width=178mm]{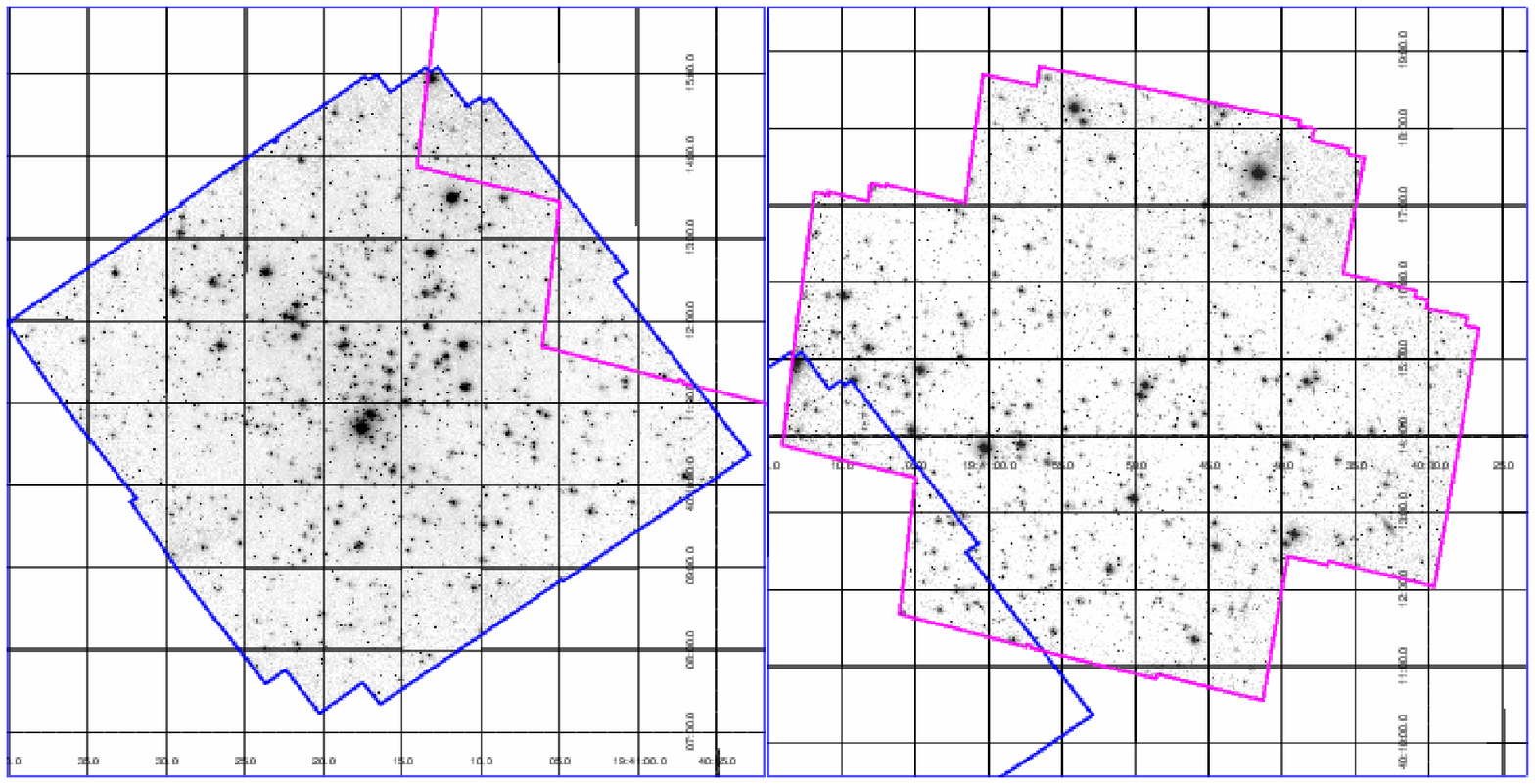}
\caption{
\textit{(Top-Right:)} Over-imposed  on the DSS is the  location of the
eight primary and secondary \textit{HST} fields employed in this work.
\textit{(Top-Left:)} The  same region as  seen in one of  the periodic
full-frame  images collected  by  the \textit{Kepler}-mission  (image:
\texttt{kplr2009114174833\_ffi-cal\_c50.fits}.
\textit{(Bottom:)}  Stacked images  of  deep exposures  in the  filter
F814W.  On the left we display the four primary WFC3/UVIS fields (blue
outline),  on the  right  the four  parallel  ASC/WFC fields  (magenta
outline).   On  both panels  we  show  reference  grids of  equatorial
coordinates  at  J2000.0.  These  images are  released  in  electronic
format, as part of this work.
\label{f1}
}
\end{center}
\end{figure*}

For the first epoch (GO-11688)  8 orbits were allocated in two filters
during October 2009, while the  second epoch (4 orbits) was in October
2012 and used only the redder of the two filters.
As primary  instrument the  Ultraviolet-Visible (UVIS) channel  of the
Wide Field Camera  3 (WFC3) gathered images in  four contiguous fields
(each 162$^{\prime\prime}$$\times$162$^{\prime\prime}$) organized in a
2$\times$2 array centered on the core of NGC\,6819.
The same  number of  fields were also  observed in parallel,  with the
Wide  Field Channel  (WFC) of  the Advanced  Camera for  Surveys (ACS)
(each   202$^{\prime\prime}$$\times$202$^{\prime\prime}$),   which  is
located  in the \textit{HST\/}  focal plane  at about  $6^\prime$ from
UVIS, and with  the detector axes oriented at  $\sim45^\circ$ from the
WFC3 axes.
Thus the primary  plus parallel exposures covered a  total of about 70
arcmin square in NGC\,6819 (see top-left panel of Fig.~\ref{f1}).

\begin{figure*}
\begin{center}
\includegraphics[width=178mm]{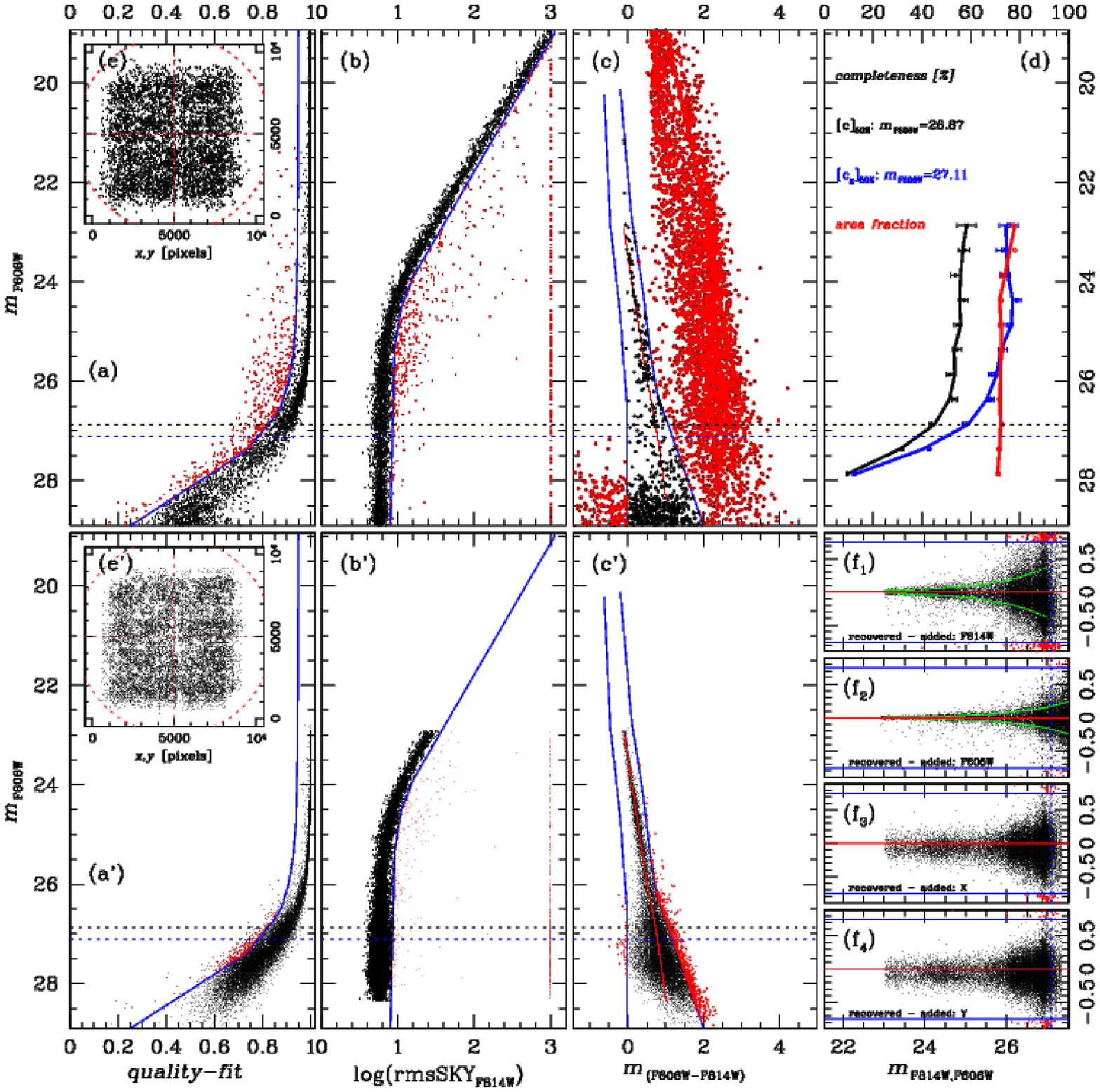}
\caption{
(a)  and  (a$^\prime$):\   The  parameter  \textit{quality-fit}  as  a
  function  of magnitude,  for real  stars and  ASs.  Objects  that we
  consider to  be real stars lie to  the right of the  blue lines. The
  discarded sources are highlighted in red.  [In all cases here except
    (d), the  blue lines  were drawn  by hand in  the lower  panel and
    copied into the upper panel.]
(b) and (b$^\prime$):\ The  parameter \textit{rmsSKY} as a function of
  magnitude.  Stars  that lie on an acceptable  background are located
  to the left of the blue lines.
(c)  and (c$^\prime$):\  The  CMD,  with blue  lines  bounding the  WD
  region.  The  red lines show the  fiducial line along  which the ASs
  were added.
(d):\ Black points and line show the traditional ``completeness'' that
  ignores  the \textit{rmsSKY}  criterion, while  the blue  points and
  line show the  completeness in the regions where  faint stars can be
  detected  and  measured.  The  red  line  shows  the  size  of  this
  measurable area in terms of fraction of total area.
(e)  and  (e$^\prime$):\  Spatial  distribution  of the  real  and  AS
  detections.   The assumed  cluster center  is indicated  by  the red
  dashed-lines. 
Panels  (f$_{1,2,3,4}$), display  the  recovered \textit{minus}  added
residual  biases in  magnitudes and  positions. The  red line  marks a
difference  equal to  zero, while  the  blue lines  mark the  interval
within which  a star is  considered recovered. The green  line denotes
the 68.27th percentile of the distributions.
\label{f2}
}
\end{center}
\end{figure*}

All collected images  were taken in the filters  F606W \& F814W, which
are the  optimal choice for our  scientific goals, i.e.,  the study of
the faintest WDs  in a relatively-old open cluster,  and are available
for both instruments (although with different zero points and slightly
different colour responses).
Data were organized in 1-orbit  visits per filter and per field.  Each
orbit  consists  of one  10\,s  short  and 4$\times$$\sim$600\,s  long
exposures for  the primary instrument,  and 3$\times$$\sim$470\,s long
exposures for  ACS.  Within each orbit both  instruments use analogous
filters.
The  bottom panels  of  Fig.\,\ref{f1}  show a  stacked  image of  the
regions, after removal of cosmic rays and most of the artifacts.

All images  were treated  with the procedures  described in  detail by
Anderson  \&  Bedin  (2010),  to  correct  positions  and  fluxes  for
imperfections in the charge-transfer  efficiency (CTE) of both ACS/WFC
and  WFC3/UVIS. Finally,  in all  images,  we masked-out  by hand  the
ghosts of the brightest stars in the fields.\footnote{
(from  WFC3 instrument handbook,  Sect.~6.5.3): \textit{The  WFC3 UVIS
    channel exhibits three different types of optical ghosts: a) those
    due  to reflections  between the  CCD  front surface  and the  two
    detector package windows; b)  those due to reflections between the
    window  surfaces;  and c)  those  due  to  reflections within  the
    particular filter in use.}
}

Photometry  and relative  positions  were obtained  with the  software
tools described  by Anderson et  al.\ (2008).  In addition  to solving
for positions  and fluxes, we  also computed two  important diagnostic
parameters:    \textit{quality-fit}    and    \textit{rmsSKY}.     The
\textit{quality-fit} essentially tells  how well the flux distribution
resembles the  shape of  the PSF (defined  in Anderson et  al.\ 2008);
this  parameter can  be very  useful  for eliminating  the faint  blue
galaxies  that plague  studies  of WDs  (see  Bedin et  al.\ 2009  for
detailed  discussions).  The sky-smoothness  parameter \textit{rmsSKY}
is the rms deviation from the  mean, for the pixels in an annulus from
3.5  to  8  pixels  from  each  source.   As  discussed  in  Bedin  et
al.\  (2008a),  \textit{rmsSKY}  is  invaluable in  measuring  a  more
effective completeness than has been used in most previous studies.

The  photometry was calibrated  to the  Vega-mag system  following the
procedures given in Bedin et  al.\ (2005b), and using the most updated
encircled energy curves and zero points\footnote{
\texttt{
http://www.stsci.edu/hst/wfc3/phot\_zp\_lbn; 
http://www.stsci.edu\-/hst\-/acs\-/analysis\-/zeropoints. 
}
}.  
In  the following  we will  use  for these  calibrated magnitudes  the
symbols $m_{\rm F606W}$ and  $m_{\rm F814W}$. The absolute accuracy of
the calibration should be good to about 0.02 magnitude per filter.

The ACS/WFC parallel fields, although useful to derive the present-day
mass function in the outskirts of  the clusters from the MS stars, are
not  very useful  to  study  the WD  CS,  as the  handful  of WDs  are
outnumbered by the field stars and background galaxies. There are also
fewer ACS/WFC  images than WFC3/UVIS  images, and their  larger pixels
size  complicates the  identification of  true point-sources,  and the
measurement of proper motions.  Also, there are no short exposures for
the WFC  images, making  these data useless  for studying  the evolved
members of NGC\,6819.
Nevertheless, for completeness and given the interest in this patch of
sky,  which falls  inside the  \textit{Kepler} field,  we  release the
catalogue   and  atlases  for   the  ACS/WFC   fields  as   well  (see
Sect.\,\ref{EM}).

Hereafter,  we   will  use  only  the  WFC3/UVIS   central  field.  In
particular, the  long exposures of the  primary field will  be used to
study the WD CS of NGC\,6819,  while the short exposures will allow us
to study the brightest and evolved cluster members.
The  photometry obtained from  the short  exposures was  corrected for
differential reddening as in Milone et al.\ (2012, see their Sect.~3),
and  carefully registered  to the  zero points  of the  long exposures
using unsaturated  common stars.  A precision of  $\sim$0.01 magnitude
per  filter was  estimated for  this  operation; this  means that  the
photometry from short and long  exposures should be consistent to this
level.

Artificial-star  (AS)  tests   were  performed  using  the  procedures
described by Anderson et al.\ (2008).  We chose to cover the magnitude
range $\sim$$23  < m_{\rm F606W}  < \, \sim$$28.5$, with  colours that
placed the artificial stars on  the WD sequence. These tests played an
important role  in showing us  what selection criteria should  be used
for the  real stars. The quality  of our results depends  very much on
making  a  good  selection,  the   details  for  which  are  shown  in
Fig.\,\ref{f2}.   (See  Bedin  et  al.\  2009,  2010  for  a  detailed
description of the selection procedures.)  Panels with unprimed labels
refer  to real  stars, while  those with  primed labels  refer  to the
artificial stars.

We  used the  AS  tests to  show  what combinations  of magnitude  and
\textit{quality-fit}  are  acceptable  for  valid star  images  [panel
  (a$^\prime$)], and we drew the  blue lines to isolate the acceptable
region.  We  then drew identical lines  in panel (a),  to separate the
real stars from blends and probable galaxies.
We  went  through  similar  steps for  the  \textit{rmsSKY}  parameter
(panels b$^\prime$ and b).
Finally we  plotted CMDs and drew  dividing lines in a  similar way to
isolate  the  white  dwarfs  (panels  c$^\prime$  and  c)  from  field
objects. Thankfully, WDs occupy  a relatively uncontaminated region of
the CMD (Bedin et al.\ 2009).

As one can see from panels  (a) and (a$^\prime$), the shape (i.e., the
\textit{quality-fit})  of the  objects  becomes increasingly  confused
with the noise at magnitudes fainter than $m_{\rm F606W}$$\sim$26, and
therefore a few blue compact  background galaxies might have fallen in
the  list  of  selected  objects.  However, these  cannot  affect  the
location of the drop of WD  star counts, which is the relevant feature
studied  in our work.   Furthermore, the  proper motions  described in
Sect.\,4, confirm the WD LF cut-off magnitude.

Our final step  was to deal with completeness,  a concept that appears
in  two different contexts:  (1) The  observed number  of WDs  must be
corrected  for  a   magnitude-dependent  incompleteness.   (2)  It  is
customary to choose the limiting reliable magnitude at the level where
the sample is 50\% complete.  In a star cluster these two aims need to
be  treated in  quite  different ways.   (1$^\prime$)  To correct  for
incompleteness the WD LF, we use the traditional ratio of AS recovered
to the number inserted.  (2$^\prime$)  For the limit to which measures
of  faint stars  are  reliable, however,  the 50\%-completeness  level
needs to be chosen in a quite different way, because in a crowded star
cluster  more faint  stars  are lost  in  the brightness  fluctuations
around the bright stars, than are  lost to the fluctuations of the sky
background.  The  completeness measure that  we should use to  set the
50\% limit is therefore the recovered fraction of ASs {\it only} among
those ASs whose value of  \textit{rmsSKY}, as a function of magnitude,
indicates that it {\it was possible to recover the star\/}.
[For  a   more  detailed  discussion,   see  Sect.\  4  of   Bedin  et
  al.\  (2008a), and for  other two  similar applications  and further
  discussions see Bedin et al.\ 2009, 2010].

The distinction  between the two completeness  measures is illustrated
in panel (d) of Fig.\,\ref{f2}, where the black line shows the overall
completeness, while  the blue  line shows the  completeness statistics
that take \textit{rmsSKY} into account; its 50\% completeness level is
at $m_{\rm F606W}=27.11$, rather  than 26.87\,mag with the traditional
statistics.   To  emphasize  the  contrast,  the red  line  shows  the
fraction of the area (i.e.,  with low-background) in which faint stars
could have been found.

%
\section{Comparison with Theory}
%

As in  our previous  similar works on  other clusters,  the comparison
with theory  comprises the determination of  a MS TO age,  and the age
implied by the WD LF.

BaSTI\footnote{\
\texttt{http://www.oa-teramo.inaf.it/BASTI}
}
scaled-solar isochrones for both WD  (Salaris et al.\ 2010) and pre-WD
(Pietrinferni  et   al.\  2004)  evolutionary   phases,  including  MS
convective-core overshooting,  were used to determine  both MS\,TO and
WD ages. We considered the  available BaSTI (pre-WD and WD) isochrones
for  [Fe/H]=+0.06, consistent with  [Fe/H]=$+$0.09$\pm$0.03 determined
spectroscopically for a sample of cluster red clump stars by Bragaglia
et al.\  (2001 -- but see  Anthony-Twarog et al.\ 2014  for a slightly
different    estimates,    [Fe/H]=$-$0.06$\pm$0.04,   obtained    from
$uvby\,Ca\,H_\beta$ observations).

First  we  determined  the  cluster distance  modulus  from  isochrone
fitting to the MS. Given that stars above the MS\,TO are saturated, we
could not  use the magnitude  of the red  clump and the colour  of red
giant branch  stars as constraints.  We have considered  the unevolved
cluster MS between $m_{\rm  F606W}$=17 (about two magnitudes below the
MS\,TO) and 19, and derived a  fiducial line by taking the mode of the
$(m_{\rm  F606W}-m_{\rm F814W})$  colour distribution  of MS  stars in
0.25\,mag wide $m_{\rm F606W}$ bins.  Before performing a least square
fitting  of the  isochrone  MSs to  the  fiducial line,  we fixed  the
cluster  $E(B-V)$ to values  between 0.10  and 0.18  mag, in  steps of
0.01~mag.
We  explored  this  range  following  the  estimate  by  Bragaglia  et
al.\ (2001),  who used temperatures  of three red clump  stars derived
from line excitation (a  reddening-free parameter) and the appropriate
theoretical  spectra, to  determine their  intrinsic  $(B-V)$ colours.
Comparison with  the observed colours  provided $E(B-V)$=0.14$\pm$0.04
(Anthony-Twarog et al.\  2014 determined $E(B-V)$=0.160$\pm$0.007 from
$uvby\,Ca\,H_\beta$   observations,  within   the  range   spanned  by
Bragaglia et al.\ estimate).
These  $E(B-V)$ values  between  0.10 and  0.18  magnitudes were  then
transformed  into  $E(m_{\rm  F606W}-m_{\rm  F814W})$  and  $A_{m_{\rm
    F606W}}$  as  described by  Bedin  et  al.\  (2009), and  assuming
$R_V$=3.1.   For  each  $E(m_{\rm  F606W}-m_{\rm F814W})$  we  finally
fitted isochrones of various ages to the fiducial line.
The procedure  worked as follows. For  a given choice  of $E(B-V)$ the
vertical  shift required  to fit  the  isochrones to  the observed  MS
fiducial  line  provided  the   apparent  distance  modulus.  This  is
unaffected  by the  isochrone  age,  given that  we  were fitting  the
unevolved MS.  Once derived the  apparent distance modulus for a given
$E(B-V)$, we compared  magnitude and shape of the  isochrones' TO with
the  observed CMD,  for ages  varying between  1.0 and  4.0~Gyr.  This
comparison allowed us to both estimate  the TO age and narrow down the
range of reddenings consistent with the isochrones.
It  turned out  that  for  some $E(B-V)$  values  (and the  associated
distance moduli), around the TO region the shape of the isochrone that
matched the observed TO magnitude was very different from the shape of
the CMD.  This implied  that the  isochrone was too  old or  too young
compared to the real cluster age, so that the TO magnitude had matched
only  on account  of an  inconsistent distance  modulus, and  hence an
inconsistent reddening.
Our isochrone fitting provided 
$E(B-V)=0.17\pm0.01$, 
$(m-M)_0=11.88\pm0.11$ [corresponding to 
a distance of 2377\,pc, and to 
$(m-M)_V=12.41\pm0.12$] and a MS\,TO age 
$t_{\rm MS\,TO}=2.25\pm0.30$\,Gyr. 
The total  age range was fixed  by the youngest  and oldest isochrones
that bracketed  the TO region  in the observed CMD  for, respectively,
the shortest and largest distance moduli obtained from the MS fitting.
The  error  on  the distance  modulus  is  an  internal error  of  our
isochrone MS-fitting  method that combines in quadrature  the error of
the least square  fit at fixed reddening, and the  effect of the error
on the reddening on the least square fit value.

The reliability of our derived distance can be assessed by comparisons
with  completely  independent   methods.   Sandquist  et  al.\  (2013)
obtained
$(m-M)_V=12.39\pm0.08$   [employing  $E(B-V)=0.12\pm0.01$]   from  the
cluster eclipsing  binary WOCS~23009,  whilst Jeffries et  al.\ (2013)
determined $(m-M)_V=12.44\pm0.07$ for the eclipsing binary WOCS~40007.
Wu,  Li \&  Hekker~(2014) determined  $(m-M)_0=11.88\pm0.14$ [assuming
  $E(B-V)=0.14\pm0.04$]  through  the  global  oscillation  parameters
${\rm \Delta  \nu}$ and ${\rm  \nu_{max}}$ and $V$-band  photometry of
the cluster red giants. 

Our isochrone-based distance estimate  is in good agreement with these
independent results based on completely different methods.

Regarding  the   age,  our  estimate  from  the   MS\,TO  agrees  with
$t=2.2\pm0.2$\,Gyr obtained  by Jeffries  et al.\ (2013)  by comparing
the  MS mass-radius  relationship of  the same  BaSTI  isochrones used
here, with the  masses and radii determined for  the components of the
cluster eclipsing binary WOCS~23009.

\begin{figure}
\begin{center}
\includegraphics[width=89mm]{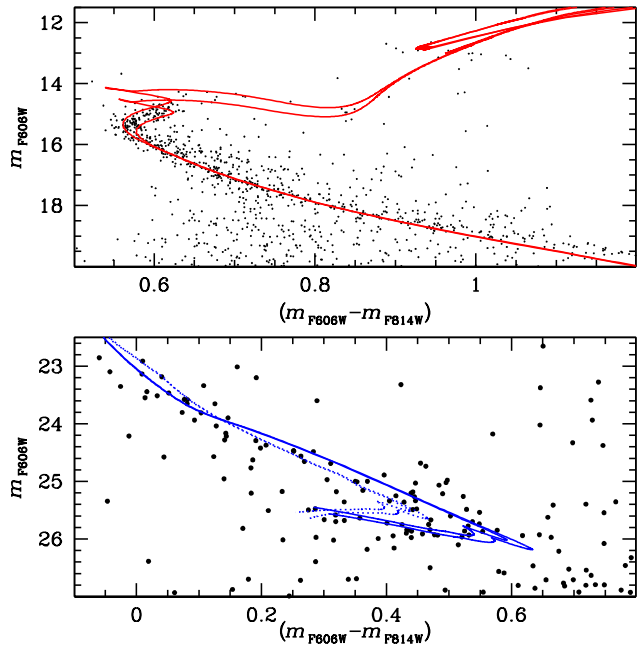}
\caption{
\textit{(Upper panel:)} 
Fit    of   theoretical    isochrones    [1.95~Gyr,   $(m-M)_0=11.99$,
  $E(B-V)$=0.18 and  2.55~Gyr, $(m-M)_0=11.77$, $E(B-V)$=0.16]  to the
cluster MS  in the $m_{\rm  F606W}$-$(m_{\rm F606W} -  m_{\rm F814W})$
CMD (see text for details).
\textit{(Lower panel:)} 
Fit   of  WD   isochrones   to   the  observed   CS,   for  the   same
age-distance-reddening  combinations.   Solid  lines  denote  DA  WDs,
dotted lines DB objects.
\label{thCMDs}
}
\end{center}
\end{figure}

\begin{figure}
\begin{center}
\includegraphics[width=89mm]{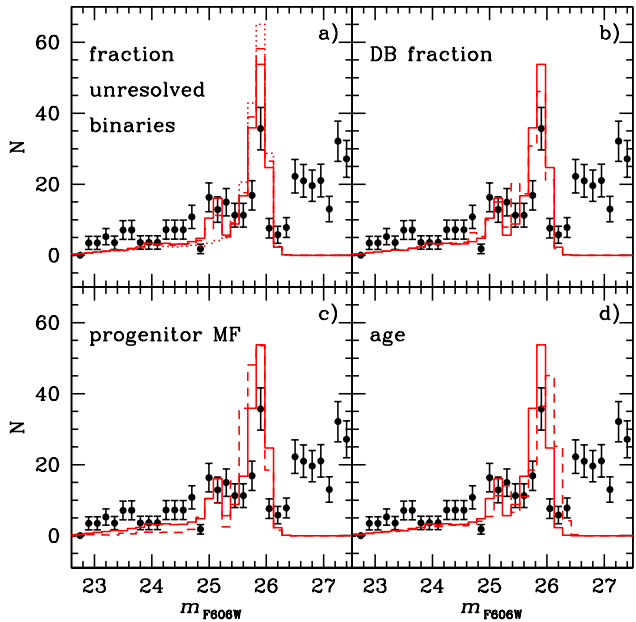}
\caption{
Comparison of  the completeness-corrected observed  $m_{\rm F606W}$ WD
LF (filled  circles with error  bars) with theoretical  2.25\,Gyr LFs,
calculated with a Salpeter MF and different choices of parameters, for
$E(B-V)=0.17$ and $(m-M)_0=11.88$.
Panel a)  displays DA  LFs with no  unresolved WD+WD  binaries (dotted
line), with a 22\% fraction  of progenitor binaries (solid line -- our
reference case), and with the same fraction of progenitor binaries but
a different distribution of mass  ratios of the two components (dashed
line -- see text for details).
Panel b) displays our reference  theoretical LF (solid line), and a LF
with 85\%  DA and 15\%  DB objects, leaving everything  else unchanged
(dashed line).
Panel c) displays  the reference LF and one calculated  with a flat MF
(dashed line -- see text for details).
Panel d) displays our reference LF (solid line) and a 2.5\,Gyr old LF,
everything else being the same (dashed line).
\label{thLFs}
}
\end{center}
\end{figure}

The upper panel of Fig.~\ref{thCMDs} shows the result of the isochrone
fits, with  the two  isochrones that best  bracket the  MS\,TO region,
i.e., 1.95\,Gyr [for  $(m-M)_0=11.99$ and $E(B-V)$=0.18] and 2.55\,Gyr
[for $(m-M)_0=11.77$ and  $E(B-V)$=0.16], respectively.  WD isochrones
(for  both DA  and DB  objects) with  the  same age-distance-reddening
combinations   are  displayed  in   the  lower   panel  of   the  same
figure. Notice how the termination of the DB sequence is brighter than
that  for the  DA WDs,  contrary to  the result  for  typical globular
cluster ages,  because in this  regime He-envelope WDs cool  down more
slowly  that  the  H-envelope  counterparts  (see,  e.g.,  Salaris  et
al.\ 2010).
To compare WD and MS\,TO  ages we have used the completeness corrected
WD LF, and matched with theoretical WD LFs of varying age the observed
cut-off of the  star counts.  The observed LF  --see points with error
bars in  Fig.~\ref{thLFs}-- exhibits just  one peak and a  cut-off (at
$m_{\rm F606W}$=26.050$\pm$0.0.075 at its faint end, as expected for a
standard cluster  CS.  Theoretical LFs  have been calculated  with the
Monte  Carlo  (MC)  technique  described  in Bedin  et  al.\  (2008b),
employing  the  BaSTI  WD models  for  both  DA  and DB  objects,  and
considering a Salpeter mass function (MF) for the progenitors.  Our MC
calculations  account for  the  photometric error  as  found from  the
data-reduction procedure  and arbitrary fractions  of unresolved WD+WD
binaries and DA and DB WDs.   The effect of random fluctuations of the
synthetic LF are minimized by using $\sim$100 times as many WDs as the
observed  sequence  (which  comprises  about  200  objects  after  the
completeness correction).

To give some more details, the MC code calculates the synthetic CMD of
a WD  population of  fixed age (and  initial chemical  composition) by
selecting first a value of the  progenitor mass for a generic WD along
the corresponding  WD isochrone (we  assume a burst of  star formation
with negligible  age and metallicity dispersion,  appropriate for open
clusters),  according to an  IMF that  is set  to Salpeter  as default
choice. The  corresponding mass and magnitudes of  the {\sl synthetic}
WD are then determined by quadratic interpolation along the isochrone,
that  has been calculated  assuming a  WD initial-final  mass relation
(see   Salaris    et   al.\   2010   for   details    about   the   WD
isochrones)\footnote{
WD isochrone tables  for a given age and  initial chemical composition
provide the mass of the  WD progenitor, the corresponding WD mass, and
the WD magnitudes in the chosen photometric filters along the CS.
}. 
For the fraction  of objects assumed to be  in unresolved binaries, we
extract randomly  also a value for  the ratio $q$  between the initial
mass of the companion and the  mass of the WD progenitor, according to
a  specified statistical  distribution.  If  the initial  mass  of the
companion has produced  a WD, we determine its  mass and magnitudes as
for single WDs.  In case the companion is  in a different evolutionary
stage, we  determine its magnitudes by interpolating  along the pre-WD
isochrone.  The fluxes  of the two components are  then added, and the
total magnitudes and colours of the composite system are computed.

We finally add  the distance modulus and extinction  to the magnitudes
of both single and unresolved binary stars, and perturbe them randomly
by using a  Gaussian photometric error with the  $\sigma$ derived from
the artificial-star tests.

A theoretical LF  is then computed for the  synthetic population to be
compared with the observed one.

In general, the  exact shape of the  WD LF (and of the  WD sequence in
the CMD) depends on a number  of parameters in addition to the cluster
age, e.g.,  the initial-final-mass relationship, the  mass function of
the progenitors,  the relative  fraction of the  (dominant) DA  and DB
objects, the fraction of  WD+WD binaries (and their mass distribution)
and, very  importantly, the dynamical evolution of  the cluster, which
selectively depletes the WDs according to their mass and their time of
formation.
Also, in  open clusters, if a  WD receives a  sufficient velocity kick
from asymmetric mass  loss during its pre-WD evolution,  it may become
unbound (see, e.g., the introduction in Williams~2004).

On one hand, this is why  we are not attempting to match precisely the
shape of  the whole WD LF,  by trying to  determine simultaneously the
cluster age, the  precise fraction of WD+WD binaries,  DB objects, and
the progenitor MF.
These effects are difficult to disentangle using just the observed LF,
and  complicated by the  limited statistic  (total number  of observed
WDs, compared for example with the  cases of M\,4 or NGC\,6791) and by
a potential (although mild) residual field contamination.

On the other  hand, the observed position of the  WD LF cut-off, which
is the primary WD age indicator, should have been affected very little
by any of these parameters.
In  the following  we  quantitatively demonstrate  how  all the  above
uncertainties, although potentially affecting  the shape of the WD LF,
do not alter the magnitude location of the LF cut-off.

First  of all,  the effect  of the  dynamical evolution  (see  Yang et
al.\  2013  for indications  about  the  dynamical  evolution of  this
cluster)  should be  negligible,  for the  following  reason.  The  WD
isochrones predict that  the brighter WDs should have  nearly a single
mass,  whereas the  region  of the  WD  LF cut-off  ---  i.e., at  the
blue-turn at the  bottom of the CS --- is expected  to be populated by
objects spanning  almost the whole  predicted WD mass  spectrum (hence
the  presence  of  a  blue  turn  towards  lower  stellar  radii).   A
mass-dependent  loss  mechanism   due  to  dynamical  evolution  would
therefore alter  the number distribution of WDs  around the blue-turn,
but it  should have no  major effect on  the observed location  of the
cut-off in the WD LF.
Figure~\ref{thLFs}   compares   the  completeness-corrected   observed
$m_{\rm  F606W}$   WD  LF  (filled  circles  with   error  bars)  with
theoretical LFs calculated to explore some of the other effects listed
above.

Panel a) displays the theoretical LF  for DA WDs (with the same number
of  objects  as  the  observed  candidate WDs),  an  age  of  2.25~Gyr
[$(m-M)_0$=11.88  and $E(B-V)$=0.17], and  two different  fractions of
progenitor binaries -- no  binaries, and 22\% respectively (dotted and
solid lines  in the  figure)-- corresponding to  zero, and  14\% WD+WD
binary  (supposed  to be  unresolved)  systems  on  the final  cooling
sequence. The  mass ratio $q$  for the binary companions  (the primary
components are assumed to follow a Salpeter MF) was chosen with a flat
distribution between 0.5 and 1.0.
As discussed,  e.g., in Podsiadlowski~(2014), the  distribution of $q$
in binary systems is not well  determined and appears to depend on the
mass  range.   Massive  binaries  favour  stars   of  comparable  mass
($q$$\approx$1) whilst the situation is less clear for low-mass stars,
and  most studies show  that $q$  is possibly  consistent with  a flat
distribution.
The 22\% fraction  of progenitor binaries was chosen  after the recent
results by Milliman et al.\  (2014), who determined a present fraction
of $22\pm3$\% MS binaries with period less than $10^4$ days.

The unresolved systems show up  in the LF at ${m_{\rm F606W}}$ between
25 and 25.2 (the variation of the star counts at fainter magnitudes is
a  consequence of keeping  the total  number of  objects fixed  at the
observed value)  about 0.75~mag brighter  than the LF cut-off,  as for
the  case  of NGC\,6791  (Bedin  et  al.\  2008b).  The  inclusion  of
unresolved  WD+WD  binaries  improves  the  fit of  the  LF  at  those
magnitudes  and leaves completely  unchanged the  magnitude of  the LF
cut-off.

We have also tested the effect  of changing the distribution of $q$ by
fixing the progenitor binary fraction to 22\%, but this time employing
a  flat distribution  between 0.1  and 1.0.  The resulting  LF (dashed
line) has a  slightly changed shape, the local  maximum between 25 and
25.2  is less  pronounced, but  the magnitude  of the  cut-off  is not
affected.
Note how  the few contaminant sources  in the sample  that might alter
our observed WD LF below  $m_{\rm F606W}$$\sim$26.5, do not affect the
location of the WD LF drop-off.

\begin{figure*}
\begin{center}
\includegraphics[width=178mm]{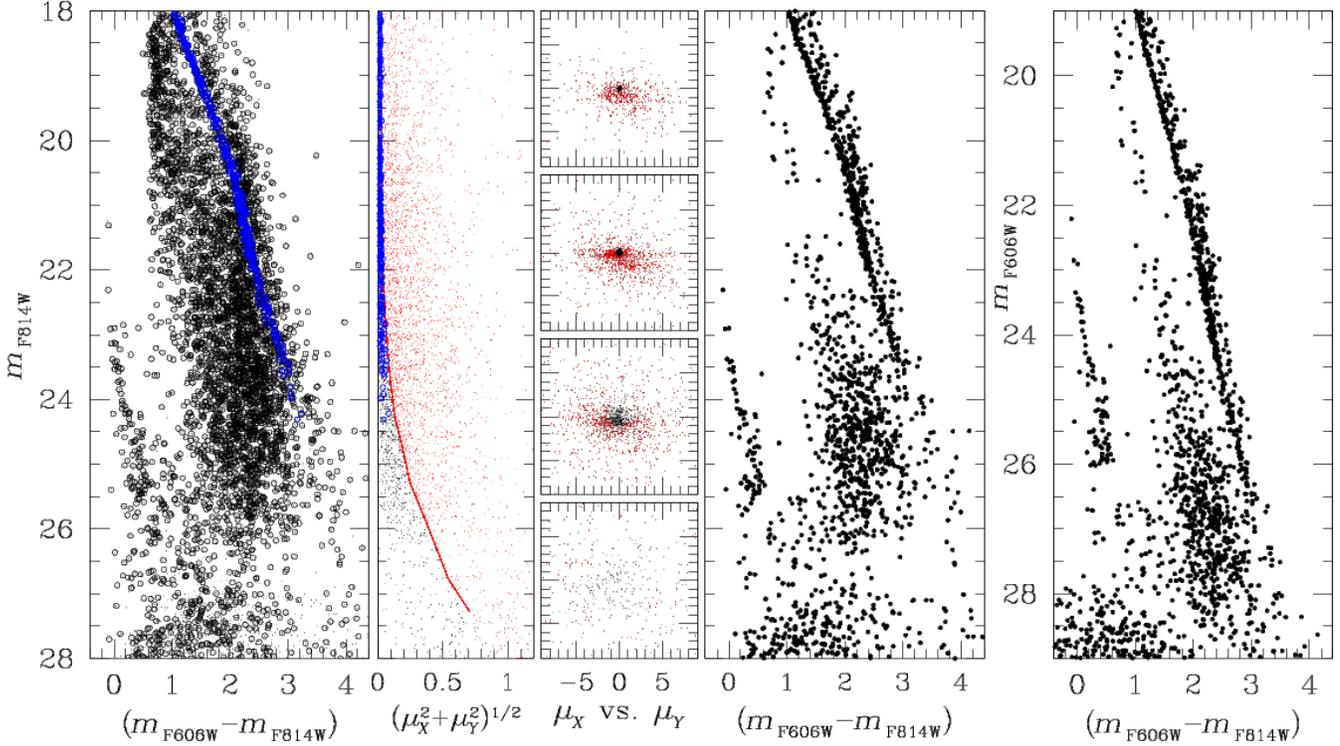}
\caption{
\textit{(Left:)} 
$(m_{\rm F606W} - m_{\rm F814W})$ vs.\ $m_{\rm F814W}$ CMD of stars in
panel (b) of  Fig.~\ref{f2} (small dots).  The stars  for which it was
possible  to  compute  a  proper-motion  are  highlighted  with  empty
circles. Blue dots  mark the MS stars used  to compute the inter-epoch
transformations necessary to compute proper motions.
\textit{(Second panel from Left:)}  
$m_{\rm F814W}$  vs.\ total proper motions  in mas\,yr$^{-1}$ relative
to  the mean  proper motion  of the  cluster.  Errors  increase toward
fainter  magnitudes.  The  red  line  isolate the  objects  that  have
member-like motions (black) from field objects (red dots).
\textit{(Third panels from Left:)} are vector point diagrams for those
stars in four magnitude bins.
\textit{(Right panels:)} resulting  decontaminated CMDs.  Note how the
sharp drop of the WD CS results even clearer than in the left panel.
\label{pms}
}
\end{center}
\end{figure*}
%

Panel b) shows  the effect of including a 15\%  fraction of DB objects
(dashed line -- DB isochrones are  also from Salaris et al.\ 2010) for
an age of  2.25~Gyr, and a reference 22\%  binary progenitor fraction.
Also in  this case the magnitude of  the LF cut off  is unchanged, and
the  signature  of DB  WDs  appears  at  $m_{\rm F606W}\sim$25.5  (the
variation  of  the  star  counts  at fainter  magnitudes  is  again  a
consequence of keeping the  total number of objects fixed).  According
to the observed LF  there is no room for a much  higher fraction of DB
objects, and actually a value close to zero would probably be a better
match to the observations.

Panel c)  investigates the combined  effect of varying  the progenitor
initial  MF plus  the cluster  dynamical evolution  that progressively
depletes the  lower mass (both  progenitors and WDs) objects.  To this
purpose we display the LF for  the standard Salpeter MF (solid line --
no DBs, 22\% of WD+WD unresolved  binaries) and a LF calculated with a
close to flat  MF for the WD progenitors (dashed  line), i.e., a power
law with  exponent $-$0.15.   Progenitor masses are  between $\sim$1.7
and  $\sim$6.5~${\rm M_{\odot}}$  at  our reference  age of  2.25~Gyr,
whilst  the final  WD masses  range between  $\sim$0.61  and 1.0~${\rm
  M_{\odot}}$ according to the  initial-final mass relation adopted by
the BaSTI WD isochrones (see Salaris et al.\ 2010 for details).

The comparison of the two LFs  shows that the magnitude of the cut-off
is unchanged, as expected. Only  the overall shape is affected and the
width of  the peak  of the  theoretical LF is  increased, but  not the
position of the  drop of the star counts beyond the  peak, that is the
age indicator.
Finally,  panel d)  displays  the effect  of  age (2.25  and 2.5  Gyr,
respectively), keeping the DB fraction to zero, and the reference 22\%
of progenitor binaries.   The older LF has a  fainter termination, and
can be  excluded by means of  the comparison with the  position of the
observed cut-off.

After  this reassessment  of the  solidity of  the LF  cut-off  as age
indicator, we have varied the cluster apparent distance modulus within
the range established  with the MS fitting and obtained  a range of WD
ages ${\rm t_{WD}}$=2.25$\pm$0.20~Gyr.
This age range has been  determined by considering all theoretical LFs
that display after the peak a sharp drop in the star counts at the bin
centred  around  $m_{\rm  F606W}$=26.05.   Notice that  the  range  is
narrower  than the MS\,TO  age range,  but completely  consistent with
${\rm t_{MS\,TO}}$.  This narrower age range
is due to the different sensitivity of TO and WD magnitude cut-off to
the population age (Salaris~2009a).

We  close our  theoretical  analysis presenting  two additional  tests
about our derived  WD ages.  First, we have  recalculated the BaSTI WD
isochrones considering  the WD  initial-final mass relation  (IFMR) by
Kalirai~(2009)  instead of  the  default IFMR  employed  by the  BaSTI
isochrones (Salaris  et al.\ 2009b).  In the WD progenitor  mass range
typical of NGC\,6819  (progenitor masses above $\sim$1.6 $M_{\odot}$),
this  different  IFMR  predicts  generally slightly  lower  progenitor
masses  at fixed WD  mass, compared  to the  BaSTI default  IFMR. This
implies that with the Kalirai et  al.\ (2009) IFMR WDs of a given mass
start to cool along the CS  at a later time, hence they reach somewhat
brighter magnitudes for  a given WD isochrone age  (we recall that the
WD isochrone age is equal to the sum of the progenitor age plus the WD
cooling  age), compared to  the case  of the  Salaris et  al.\ (2009b)
IFMR.   Comparisons with  the observed  LF show  that the  use  of the
Kalirai et  al.\ (2009) IFMR increases  the estimate of the  WD age by
just 40~Myr.

As  a second  test, we  have considered  the DA  WD cooling  tracks by
Renedo   et  al.\   (2010  --   their  calculations   for  metallicity
approximately  half-solar,  the closest  to  NGC\,6819 metallicity  in
their calculations), that are completely independent from the BaSTI WD
models employed  here. Renedo et  al.\ (2010) calculations  follow the
complete  evolution of the  WD progenitors  through the  thermal pulse
phases   and  include  the   effect  of   CO  phase   separation  upon
crystallization,  like  BaSTI WD  models.  The  CO stratification  and
envelope  thickness of Renedo  et al.\  models (as  well as  the model
input  physics) are  slightly  different from  BaSTI  models, but  the
cooling times down  to the luminosities of the  bottom of the observed
CS of NGC6819  are similar. From their calculations,  the match of the
cut-off  of the  observed LF  provide an  age older  by $\sim$100\,Myr
compared to the age determined with our reference BaSTI WD models.

%
\section{Proper Motions}
%

In  principle a  mild  contamination from  field  objects, might  also
affect  the shape  of  the WD  LF,  but again,  it  cannot change  the
magnitude location of the sharp drop off in the WD LF. In this section
we  use proper  motions to  strengthen the  cluster membership  of the
observed stars in the expected CMD location of the WD CS.

Proper motions were determined by means of the techniques described in
Bedin et al.\ (2003, 2006, 2009, 2014).  We used the deep exposures in
F814W  collected in October  2009 as  first epoch,  while we  used the
F814W exposures taken in October 2012 as second epoch.
The  proper motions  are calculated  as the  displacement  between the
average  positions  between  the  two  epochs,  divided  by  the  time
base-line (about 3 years), and  finally multiplied by the pixel scale.
We  assumed  a  WFC3/UVIS  pixel  scale  of  39.773  mas\,pixel$^{-1}$
(Bellini, Anderson \& Bedin 2011).
As the  proper motions signals are  a function of  the F814W magnitude
only,  we  plot  proper  motions  properties as  a  function  of  this
magnitude, rather than F606W.
 
Unfortunately,  the   cluster  has  a   proper  motion  that   is  not
significantly different  from the bulk  of the field objects,  and our
precision  is too  low for  an accurate  discrimination  between field
objects  and cluster  members,  especially at  the  magnitudes of  the
faintest WDs.   Furthermore, proper motions are not  available for all
sources, with an incompleteness  that is hard to assess quantitatively
(and reliably);  the precision of  the derived proper motions  is also
heterogeneous across the sample.

Even though we cannot do quantitative work with the proper motions, we
can still use them in a qualitative way to confirm:
\textit{(i)}  the truncation of the WD LF, 
\textit{(ii)} the observed shape of the end of the WD CS 
      (as the stars having proper motions are also the ones with the best photometry), and 
\textit{(iii)} the goodness of the employed isochrones (how well they reproduce the blue-turn).

The  left panel of  Fig.~\ref{pms} shows  as small  dots all  stars in
panel (c)  of Fig.~\ref{f2},  but this time  plotting the  colour {\it
  vs.}\,$m_{\rm F814W}$ instead  of {\it vs.}\,$m_{\rm F606W}$.  Stars
for which it was possible  to estimate a proper motion are highlighted
with open circles.
Only  members along  the MS  were used  to define  the transformations
between the two epochs, and they are marked in blue.
In  the  second  panel, we  show  for  this  sub-sample of  stars  the
magnitudes of the proper motions vs.\ $m_{\rm F814W}$. Cluster members
should have a dispersion of the order of $\sim$1 km s$^{-1}$, which at
a  distance of  $\sim$2.4\,kpc  would correspond  to  a proper  motion
internal dispersion  of less than 0.1\,mas\,yr$^{-1}$,  making them an
ideal almost-fixed reference frame.
For  this reason  our  zeroes of  the  motion will  coincide with  the
cluster bulk motion.

It can be seen that  at high-signal-to-noise ratio, cluster stars have
a  sharp proper  motion distribution  which is  consistent with  a few
times 0.1\,mas\,yr$^{-1}$,  but embedded in  the middle of a  cloud of
field objects, mainly thin disk  and thick disk stars [NGC\,6819 is at
  (\textit{l,b})$\simeq$(74$^\circ$,8$^\circ$)].   Note the increasing
size of random errors with decreasing signal-to-noise ratio.

In order to separate members from non-members we used the results of
the artificial star test illustrated in (f$_3$ and f$_4$) and derive
the 1-D $\sigma$ as the 68.27th percentile of the added$-$recovered
positions for each coordinate at different magnitude steps.  The red
line marks the 2.5\,$\sigma$ level.

Non-members, to the  right of the red line, are marked  in red. In the
third  panels  from  left,  using   the  same  symbols,  we  show  the
vector-point diagrams (in mas  yr$^{-1}$) for four different magnitude
bins ($m_{\rm  F606W}$ = 18-20.5, 20.5-23, 23-25.5,  and 25.5-28). The
CMDs that results from the  cleaning criterion of the second panel are
shown in the rightmost panels of the same figure, using both filters.
No matter how  tight the circle is, even  at the brightest magnitudes,
there will  always be  field objects that  accidentally have  the same
motions of NGC\,6819 members.  Thankfully, the WDs of NGC\,6819 occupy
a  low-contamination region  of the  CMD,  were few  blue faint  stars
exist,  field  WDs and  blue  compact  galaxies  both being  generally
significantly fainter (Bedin et al.\ 2009).

Although affected by  low completeness, it seems even  more clear that
the   sharp  drop   at  $m_{\rm   F606W}\simeq26.05$   (i.e.,  $m_{\rm
  F814W}\simeq25.6$)  in Fig.~\ref{f2}(c)  is  real and  that the  few
objects along the  continuation of the WD CS  at $m_{\rm F606W}\sim27$
(i.e.,  $m_{\rm   F814W}\simeq26$)  are  most  likely   field  WDs  or
background unresolved galaxies.

\begin{figure}
\begin{center}
\includegraphics[width=89mm]{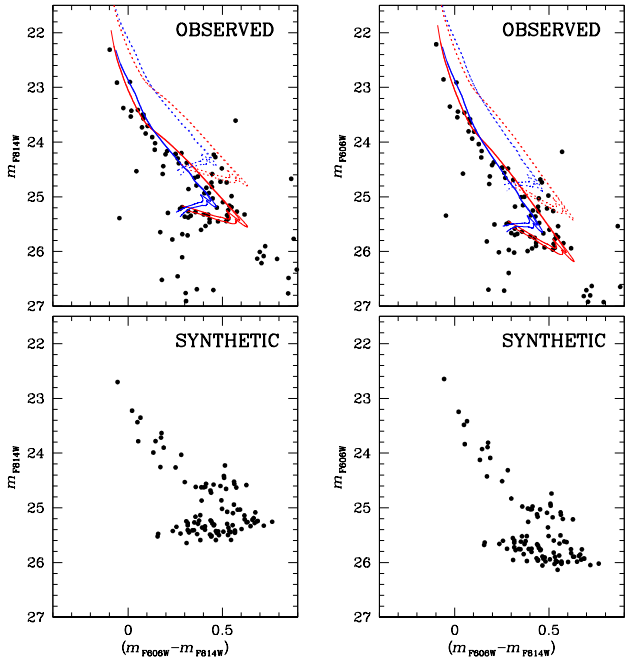}
\caption{
A zoom-in of the decontaminated  CMDs in Fig.~\ref{pms} centred on the
WD CS,  with colour  \textit{vs.} $m_{\rm F814W}$  in left  panels and
\textit{vs.}\ $m_{\rm  F606W}$ in  the  right  panel.  The top  panels
display  the  observed  CMDs,  with  over-imposed  the  WD  isochrones
employed in Fig.~\ref{thCMDs} (solid lines: DA in red, DB in blue).
We also  display, as  dotted lines, the  same isochrones  but 0.75~mag
brighter, to mark approximately the magnitude and colour range covered
by a population of unresolved WD+WD binaries.
The  bottom panels  display the  synthetic CMDs  corresponding  to the
reference LF  shown as  a solid line  in panel b)  of Fig.~\ref{thLFs}
(see Sect.~3 for details) with  approximately the same number of stars
of the decontaminated CMDs (see Sect.~3 for details).
\label{WDcln}
}
\end{center}
\end{figure}

Finally, in Fig.~\ref{WDcln} we show the CMDs of the decontaminated WD
CS. Although very incomplete, these contain the best measured WDs in
our sample, offering a higher-photometric-precision version of the
complete sample used to derive the WD LF. Indeed, the shape of the
observed blue-turn here appears to be significant, and most
importantly consistent with the shape of the WD DA isochrones
(indicated in red) and with the synthetic CMDs for the reference LF of
panel b) in Fig.~\ref{thLFs}.
Very   interestingly,   the  synthetic   CMDs   show  around   $m_{\rm
  F606W}\sim25$ and  $(m_{\rm F606W}-m_{\rm F814W})\sim$0.5  a feature
caused  by the unresolved  WD+WD binaries  included in  the simulation
(see  Sect.~3 for  details), that  corresponds to  the end  of  the DA
cooling sequence  shifted 0.75\,mag brighter (see upper  panels of the
same  figure).   The observed  CMDs  exhibit  a  very similar  feature
located at the same magnitude and colour, hinting at the presence of a
small fraction of  unresolved WD+WD systems, as suggested  also by the
fit of the LF (see Section~3).

A  future paper  (involving  other team-members)  will  deal with  the
absolute motions  of NGC\,6819 with respect to  background galaxies in
order to compute the cluster Galactic orbit using as input the sources
in the catalogue described in  the next section. [Similar to the study
 of the orbit of NGC\,6791 done in Bedin et al.\ 2006.]

%
\section{Electronic Material: WFC3/UVIS and ACS/WFC}
\label{EM}
%

\begin{figure}
\begin{center}
\includegraphics[width=89mm]{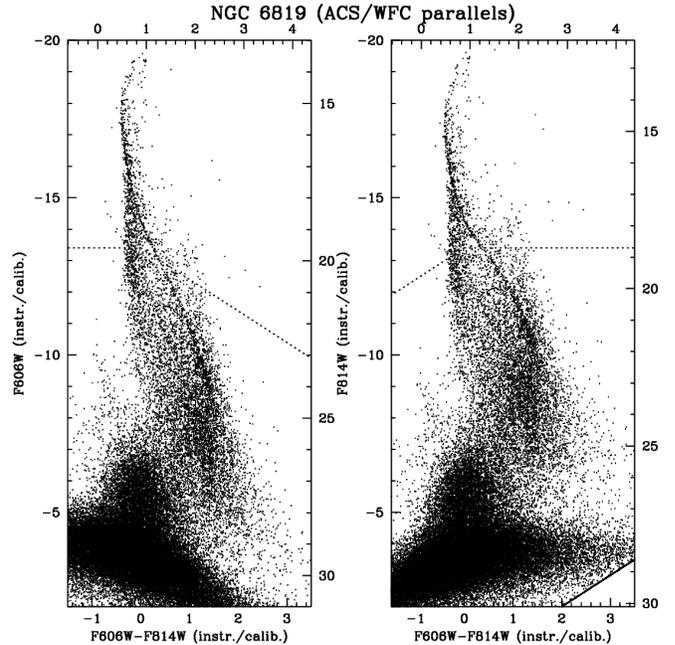}
\caption{
Colour-Magnitude  diagrams obtained from  the ACS/WFC  parallel fields
shown in the bottom-right panel of Fig.\,\ref{f1}. The bottom and left
axes  display instrumental  magnitudes, while  the top  and  right the
calibrated  ones. Dashed  lines  indicate the  level where  saturation
sets-in. Note that it is  possible to recover photometry for saturated
stars up to about 3.5 magnitude above these limits (see Gilliland 2004
for details), but not enough to observe the MS TO stars reliably.
\label{acs}
}
\end{center}
\end{figure}

The catalogue  is available electronically  in this Journal  (and also
under  request  to the  first  author).  In  the catalogue,  Col.\,(1)
contains  the  running  number;  Cols.\,(2)  and (3)  give  the  J2000
equatorial coordinates in decimal degrees for the epoch J2000.0, while
Cols.\,(4)  and  (5)  provide the  pixel  coordinates  x  and y  in  a
distortion-corrected  reference  frame.   Columns  (6)  to  (11)  give
photometric data, i.e., $m_{\rm F606W}$ and $m_{\rm F814W}$ magnitudes
and their errors. If photometry in a specific band is not available, a
flag equal  to 99.999  is set for  the magnitude  and 0.999 -  for the
error.
In  Fig.~\ref{acs}  we show  the  CMDs  obtained  from the  unselected
catalogue of the  ACS/WFC parallel fields, which we  have not analysed
and used in  this article.  For ACS/WFC we provide  the entire list of
all detections, these still include blends and other artifacts.
For the case of WFC3/UVIS main catalogue we also provide the subset of
the selected stars  as shown in Fig.\,\ref{f2} and a  WD flag. We also
provide the photometry from short exposures.

The stack images shown in  the bottom panels of Fig.\,\ref{f1} provide
a  high-resolution  representation  of  the  astronomical  scene  that
enables us to examine the region around each source.
Indeed,  the  pixel-based rapresentation  of  the  composite data  set
provides   a   nice  objective   cross-check   on  the   catalog-based
representations,  which is  a product  of our  subjective  finding and
selection criteria.
The   stacked  images   are  20\,000$\times$20\,000   for   UVIS  (and
24\,000$\times$24\,000 for  WFC) super-sampled pixels (by  a factor of
2,  i.e., 20\,mas\,pixel$^{-1}$  for  UVIS, 25\,mas\,pixel$^{-1}$  for
WFC).   We  have  included  in  the  header of  the  image,  as  World
Coordinate System keywords, our absolute astrometric solution based on
the  2MASS point source  catalogue. As  part of  the material  of this
article,  we also electronically  release these  astrometrized stacked
images.

%
\section{CONCLUSIONS}
%

We  have  used  \textit{Hubble  Space  Telescope} to  observe  the  WD
sequence  in  NGC\,6819, a  near  solar-metallicity  old open  cluster
within the $Kepler$ mission field of view.  By means of our photometry
and completeness tests we have determined the LF of the cluster WD CS,
which exhibits a sharp drop at $m_{\rm F606W} = 26.050 \pm 0.075$, and
a  shape  consistent  with  theoretical predictions  for  a  canonical
cluster CS.

Isochrone fits to the cluster MS have provided 
$E(B-V)=0.17\pm0.01$, 
$(m-M)_0=11.88\pm0.11$, and a MS\,TO age 
$t_{\rm MS\,TO}=2.25\pm0.30$\,Gyr. 
These distances and ages are in agreement with independent constraints
from eclipsing binary and asteroseismological analyses.

We have reassessed the robustness of the WD LF cutoff magnitude as age
indicator  and determined an  age of  $2.25\pm0.20$\,Gyr from  the CS,
completely consistent with the age of $2.25\pm0.30$\,Gyr obtained from
the MS\,TO.
We  have  tested the  effect  of  a different  IFMR  on  our WD  ages,
obtaining  an age  larger by  just 40~Myr  when using  the  Kalirai et
al.\ (2009) IFMR instead of our reference Salaris et al.\ (2009) IFMR.
We  have   also  employed  the   completely  independent  set   of  WD
calculations by Renedo  et al.\ (2010) and determined  an age older by
just 100\,Myr compared to the result with our reference WD models.

This agreement between WD and MS\,TO  ages is in line with the results
we obtained in  our previous studies of NGC\,2158  and M\,67, two open
clusters spanning the age range between $\sim$2 and $\sim$4~Gyr.

As a  by-product of this work  we release, in  electronic format, both
the  catalogue of  all the  detected sources  and the  atlases  of the
region (in two filters), contained in the $Kepler$ field.

\section{Acknowledgments}
J.A.\ and  I.R.K.\ acknowledge support  from STScI grant  GO-11688 and
GO-12669.  P.B.\ was supported in part  by the NSERC Canada and by the
Fund FRQ-NT (Qu\'ebec).

\noindent


%


\label{lastpage}


\end{document}